\documentclass[conference]{IEEEtran}
\ifCLASSINFOpdf
\else
\fi
\usepackage{graphicx}
\usepackage{setspace}
\usepackage{amsmath}
\usepackage{amssymb} 
\usepackage{bm} 
\usepackage{cite}
\usepackage{float}
\usepackage[usenames,dvipsnames]{pstricks}
\usepackage{epsfig}
\usepackage{pst-grad} 
\usepackage{pst-plot} 
\usepackage{tabularx}
\usepackage[keeplastbox]{flushend}
\usepackage[T1]{fontenc}
\newcommand{\lb} {\left}
\newcommand{\rb} {\right}
\newcommand{\nn} {\nonumber}

\allowdisplaybreaks
\flushend
\usepackage{caption}
\usepackage{subcaption}
\usepackage[utf8]{inputenc}

\begin{document}

\onecolumn{\noindent © 2021 IEEE. Personal use of this material is permitted. Permission from IEEE must be obtained for all other uses, in any current or future media, including reprinting/republishing this material for advertising or promotional purposes, creating new collective works, for resale or redistribution to servers or lists, or reuse of any copyrighted component of this work in other works.}
 \twocolumn{
\bstctlcite{IEEEexample:BSTcontrol}
\title{Transmitter Selection for Secrecy in a Frequency Selective Fading Channel with Unreliable Backhaul}
\author{
  \IEEEauthorblockN{Shashi Bhushan Kotwal\IEEEauthorrefmark{1}, Chinmoy Kundu\IEEEauthorrefmark{2}, 
  Sudhakar Modem\IEEEauthorrefmark{3},
  Ankit Dubey\IEEEauthorrefmark{4}, 
  and Mark F. Flanagan\IEEEauthorrefmark{5}
  }
  \IEEEauthorblockA{\IEEEauthorrefmark{1}\IEEEauthorrefmark{3}\IEEEauthorrefmark{4}Department of EE, Indian Institute of Technology Jammu, Jammu \& Kashmir, India }
  
  \IEEEauthorblockA{\IEEEauthorrefmark{2}\IEEEauthorrefmark{5}School of Electrical and Electronic Engineering, University College Dublin, Belfield, Ireland}
  \textrm{
  {{\IEEEauthorrefmark{1}sbkotwal@ieee.org},
  \{\IEEEauthorrefmark{3}sudhakar.modem, \IEEEauthorrefmark{4}ankit.dubey\}@iitjammu.ac.in}, {\IEEEauthorrefmark{2}chinmoy.kundu@ucd.ie}, 
  {\IEEEauthorrefmark{5}mark.flanagan@ieee.org}}}
\maketitle
 \pagestyle{empty}
\begin{abstract}
In this paper, a communication network using single carrier with cyclic prefix modulation over frequency selective channels is considered, where an access point provides connectivity to a legitimate destination through multiple transmitters with unreliable backhaul links in the presence of an eavesdropper. A sub-optimal and an optimal transmitter selection scheme are proposed to improve the secrecy of the system, depending on whether the active backhaul channel knowledge is available \textit{a priori} or not. The secrecy outage probability (SOP) and its asymptotic limit are presented in closed-form. This provides some insights regarding how knowledge of the active backhaul links affects the secrecy performance of the network. Our results show that the optimal transmitter selection scheme obtains a larger benefit than the sub-optimal scheme from the knowledge of the active backhaul links, resulting in a significantly improved system performance; however, the sub-optimal transmitter selection scheme can reduce the complexity and feedback overhead. 
\end{abstract}

\section{Introduction}

Data security in emerging wireless networks is critical due to the broadcast nature of the information-bearing signals.
Physical layer security (PLS) is becoming important in emerging wireless networks as it can reduce the complexity of implementing data encryption at the upper layers \cite{Leung_1978_Gaussian_Wiretap_Channel}. This may be beneficial in future networks where low-complexity nodes are deployed, e.g., internet-of-things (IoT).

It was shown in \cite{Mukherjee_2011_beamforming} that in a multiple-input single-output (MISO) antenna system, beamforming with or without perfect channel state information (CSI) can improve
the signal-to-noise ratio (SNR) at the legitimate receiver.
Beamforming in general requires additional hardware, computation, and communication overhead, which might make it less attractive for implementation in power-constrained wireless nodes in low-complexity networks \cite{Sheng_2018_beaforming_difficult}. In contrast, selecting one transmitter out of many based on perfect or imperfect CSI appears to be a simpler way to achieve a diversity benefit in future wireless networks with low-complexity terminals \cite{Chinmoy_TWC_2015}. This technique requires only CSI feedback to make the decision for the selection. Depending on the available CSI, many sub-optimal and optimal transmitter and or relay selection schemes exist in the literature \cite{Chinmoy_TWC_2015, Chinmoy_GC16,Chinmoy_GC17,Kim_2015_CPSC,other_2017_CPSC,Kim_2016_CPSC_Globecom,Kim_2016_CPSC_Trans,Kim_2018_CDD_JOUR,Chinmoy_TVT_2019,Shalini_GC20,backhaul_Yin}. From the secrecy perspective, the sub-optimal selection (SS) scheme corresponds to selecting the transmitter for which the transmission rate from the transmitter to the legitimate destination is maximized, without taking the eavesdropping channel into consideration. On the other hand, the optimal selection (OS) scheme is the one in which the secrecy rate of the system is maximized by taking global channel knowledge into consideration \cite{Chinmoy_TVT_2019}. Secrecy performance is derived mostly in terms of the secrecy outage probability (SOP), which is the probability that the secrecy rate of the system is less than a threshold value. 

Moreover, in future dense networks, the use of wireless backhaul is a more cost-effective solution than using wired backhaul \cite{Kim_2016_CPSC_Globecom}. Backhaul links are the wired/wireless connections which provide the communication links between the primary source and multiple transmitters \cite{backhaul_Yin}. Transmitter selection for secrecy along with wireless backhaul is explored in \cite{Kim_2016_CPSC_Trans,Kim_2016_CPSC_Globecom,Kim_2018_CDD_JOUR, backhaul_Yin,Chinmoy_TVT_2019,Shalini_GC20}.
The availability of backhaul activity knowledge before transmitter selection is not guaranteed in all the practical scenarios \cite{Chinmoy_TVT_2019}. There can be two possible scenarios regarding knowledge of backhaul activity. In the first scenario, the set of active backhaul links is not known  \textit{a priori} (``knowledge unavailable'' (KU) scenario). This results in selecting a transmitter irrespective of its state of activity. In the second scenario, the backhaul links which are active are known \textit{a priori} (``knowledge available'' (KA) scenario) and the selection takes place from the set of transmitters with active backhaul links. Knowledge of the backhaul activity can be acquired through the usual feedback mechanism between the access point and the transmitters.

The use of a broadband channel necessarily means encountering frequency selective fading. In frequency selective fading channels, inter-symbol interference (ISI) is a major issue and can place limitations on the achievable data rate \cite{bailleu_2020}. By applying a cyclic prefix greater than the maximum delay spread at each subcarrier, orthogonal frequency division multiplexing (OFDM) counters the effects of ISI \cite{Tubbax_SC_CP_first}. 
However controlling the peak-to-average power ratio (PAPR) in OFDM is difficult owing to the complex power amplifier design. 
The use of single carrier-cyclic prefix (SC-CP) modulation is proposed as an alternative to multi-carrier modulation \cite{Tubbax_SC_CP_first}. 
Secrecy with transmitter and/or relay selection in a frequency selective fading channel using SC-CP signaling and wireless backhaul, using sub-optimal and optimal selection schemes, has been extensively studied in \cite{Kim_2015_CPSC,Kim_2016_CPSC_Globecom,Kim_2016_CPSC_Trans,other_2017_CPSC,Kim_2018_CDD_JOUR}.

In \cite{Kim_2015_CPSC}, the authors considered a perfectly reliable wireless backhaul with multiple relays and destinations, and a cluster of eavesdroppers. The authors presented a two-stage relay and destination selection process which is sub-optimal. To degrade the channel between relay and eavesdropper, the authors in \cite{other_2017_CPSC} considered a jammer along with multiple relays. The impact of imperfect backhaul links was taken into consideration; however, a two-stage transmitter and relay selection process was considered, which is not optimal.
In \cite{Kim_2016_CPSC_Globecom}, the authors considered a relayed communication system with multiple transmitters, a single destination, and an eavesdropper. A transmitter is selected to maximize the SNR at the relay, taking into account imperfect backhaul. However, the selection was sub-optimal.
To improve the secrecy in a more severe scenario with imperfect backhaul, a similar sub-optimal transmitter selection scheme to that of \cite{Kim_2016_CPSC_Globecom} was considered in a multiple-eavesdropper scenario in \cite{Kim_2016_CPSC_Trans}. The authors in \cite{Kim_2018_CDD_JOUR} selected  the transmitter-relay pair which maximizes the end-to-end SNR of the transmitter to destination channel. However, the transmitter-relay pair selection process is sub-optimal. Thus, the existing literature in this area for the case of frequency selective channel with  SC-CP modulation considers only sub-optimal transmitter selection. However, this cannot guarantee the maximum secrecy rate. Moreover, authors of \cite{Kim_2015_CPSC,Kim_2016_CPSC_Globecom,Kim_2016_CPSC_Trans,other_2017_CPSC,Kim_2018_CDD_JOUR} considered only the KA scenario. 

Motivated by the above discussion, in this paper we propose two transmitter selection schemes, i) OS and ii) SS in the KU and KA scenarios for improving secrecy in a frequency selective fading channel with SC-CP modulation  where an access point is providing unreliable wireless backhaul connections to multiple transmitters. 
We consider a different system model to those studied in \cite{Kim_2015_CPSC,Kim_2016_CPSC_Globecom,Kim_2016_CPSC_Trans,other_2017_CPSC,Kim_2018_CDD_JOUR}; in particular, neither the OS scheme nor the KU backhaul activity scenario have been previously considered in the literature. We generalize the analysis by including the backhaul reliability parameter into the derivation of the KU as well as KA scenarios. We also present the asymptotic limit of the SOP for the OS and SS schemes in the KA and KU scenarios.  
We show that the asymptotic limit of the SOP in each case depends on the secrecy rate threshold, the number of channel paths and the path loss factor in addition to the backhaul reliability and the number of transmitters, in contrast to the system considered in \cite{Kim_2015_CPSC,Kim_2016_CPSC_Globecom,Kim_2016_CPSC_Trans,other_2017_CPSC,Kim_2018_CDD_JOUR}.

Our major contributions are listed as follows:

\begin{itemize}
\item To improve the secrecy performance over frequency selective fading channels with SC-CP modulation, we propose two transmitter selection schemes OS and SS. This is in contrast to the existing works in which only SS scheme is available \cite{Kim_2015_CPSC, Kim_2016_CPSC_Globecom, Kim_2016_CPSC_Trans, other_2017_CPSC, Kim_2018_CDD_JOUR}.
\item We generalize the analysis to integrate the backhaul reliability parameter into the SOP performance for the KU and KA scenarios, and we find the SOP expressions for the OS and SS schemes in closed-form.
\item We compare the performance of the transmitter selection schemes in the schemes in the KA and KU scenarios.
\item The secrecy performance floor is determined by deriving the asymptotic SOP for all of the aforementioned cases.
\end{itemize}

The remainder of the paper is organised as follows. Section II describes the system model, Section III analyses the secrecy outage probability in the KU scenario, while Section IV performs the corresponding analyses for the KA scenario. Asymptotic analysis is performed in Section V. Numerical results are presented in Section VI, and conclusions are presented in Section VII.

\textit{Notation:} The probability of occurrence of an event is represented by $\mathbb{P}[\cdot]$, 
$F_{X} (\cdot)$ represents the cumulative distribution function (CDF) of a random variable (RV) $X$, and $f_{X} (\cdot)$ is the corresponding probability density function (PDF).

\section{System Model}
We consider a system where an access point $A$ is providing unreliable wireless backhaul links $b_k$ to $K$ transmitters $S_k$, $k\in\{1,2,\ldots,K\}$, communicating with a single legitimate destination, D in the presence of an eavesdropper, E.
Each transmitter is equipped with one antenna, considering the limitations of deploying multi-antenna systems in future dense heterogeneous networks. 
A frequency selective fading model is considered between all transmitters and receivers such that the destination channel $\text{S}_k-\text{D}$ and the eavesdropping channel $\text{S}_k-\text{E}$, respectively, are modelled as independent and identically distributed complex Gaussian random vectors $\boldsymbol{h}_{Dk}$ and $\boldsymbol{h}_{Ek}$ with $M$ and $N$ elements, respectively, for each $k$. $M$ and $N$ denote the number of multipath components that each transmitter will encounter to D and E, respectively.

Accordingly, the distribution of each entry is given as $h_{Dk}(i)\sim \mathcal{CN}(0,1)$ and $h_{Ek}(j)\sim \mathcal{CN}(0,1)$ where $i \in\{1,2,\dots,M\}$ and $j \in \{1,2,\dots,N\}$ are circularly symmetric complex Gaussian random variables. The path loss factors for D and E, represented by $a$ and $b$ respectively, are not identical. As in \cite{Chinmoy_TVT_2019, Kim_2018_CDD_JOUR}, the reliability of the $k$th backhaul link from $A$ to $\text{S}_k$ is modeled as a Bernoulli RV $\mathbb{I}_k$ with probability of active link and inactive link indicated by $\mathbb{P}\left[\mathbb{I}_k =1\right]=\zeta$ and $\mathbb{P}\left[\mathbb{I}_k =0\right]=(1-\zeta)$, respectively. The noise contributions at D and E are modeled as independent circularly symmetric complex Gaussian random variables with mean zero and variance $\sigma^2$. 
\begin{figure}\label{system_model}
 \centering
 \includegraphics[width=4cm]{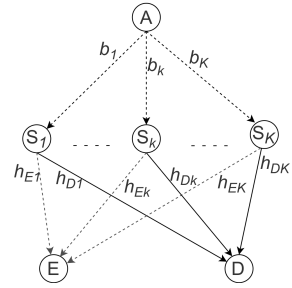}
 \vspace{0cm}
 \caption{System model. }
 \vspace{-.6cm}
\end{figure}
As a result of the SC-CP transmission considered, the received SNR at D for $S_k$ can be written when $b_k$ is active as \cite{Kim_2015_CPSC} \begin{align}
\gamma_{Dk} = A_{D} ||\boldsymbol{h}_{Dk}||^2 = \frac{ a P_T }{\sigma^2} ||\boldsymbol{h}_{Dk}||^2, 
\end{align}
where $P_T$ is the transmitted power. The path loss factor $a$ is inversely proportional to the distance of D from $\text{S}_k$. The PDF of the received SNR follows chi-squared
distribution with $2M$ degrees of freedom and can be expressed as \cite{Kim_2015_CPSC}, 
\begin{align}
\label{pdf_gen}
 f_{\gamma_{Dk}}(x) = \frac{{x}^{M-1} e^{-\frac{x}{A_D}}}{\Gamma(M)(A_D)^M},
\end{align}
where $\Gamma(\cdot)$ denotes Gamma function. It may be noted that
the distribution in (\ref{pdf_gen}) is a Gamma distribution with integer parameter $M$. This shows that when SC-CP modulation is applied in a frequency selective (multipath) fading channel where each of the $M$ paths follows an independent Rayleigh fading distribution, the resulting channel behaves as a narrowband Nakagami-m channel.
The CDF of the received SNR at D corresponding to the $k^{th}$ transmitter is
\begin{align}
\label{CDF_Nakagami}
 F_{\gamma_{Dk}}(x) 
 &=\int_0^x \frac{{y}^{M-1} e^{-\frac{y}{A_D}}}{\Gamma(M)(A_D)^M}dy = \frac{\gamma(M,x/A_D)}{\Gamma(M)},
\end{align}
where $\gamma(u,v) = \int_0^vt^{u-1}e^{-t}dt$ is the lower incomplete Gamma function. 
Similarly, $\gamma_{Ek}$ and its distribution can be expressed simply by replacing D with E, $M$ with $N$, and $a$ with $b$. When $b_k$ is inactive, the received SNR is equal to zero.
We will propose SS and OS schemes for both the KU and KA scenarios, capturing different levels of prior knowledge of the active backhaul links.
These transmitter selection schemes are executed at the access point or at a central processing unit, to which all of the necessary CSI is fed back.

The SOP of the system is defined as the probability that the secrecy rate $C_S$ of the system is below a certain required threshold rate $R_{\text{th}}$. The SOP can be expressed, including backhaul uncertainty and transmitter selection, as \cite{Chinmoy_TWC_2015}
\begin{align}\label{SOP_SUBOP_BH_Unknown1_1}
 P_{\text{out}}{(R_{\text{th}})}&=\mathbb{P}[C_S < R_{\text{th}}] = \mathbb{P}\lb[\frac{1+\hat{\gamma}_{D^*}}{1+\hat{\gamma}_{E^*}} < \rho\rb] \nonumber \\
 &=\int_{0}^{\infty} F_{\hat{\gamma}_{D^*}}(\lambda) f_{\hat{\gamma}_{E^*}}(y)dy,
\end{align}
where $\rho=2^{R_{\text{th}}}$, $\lambda=((1+y)\rho-1)$, $\hat{\gamma}_{D^*}$ and $\hat{\gamma}_{E^*}$ are the SNR at D and E, respectively, corresponding to the selected transmitter $k^*$ including backhaul uncertainty. As the eavesdropping channel SNR improves, the SOP increases.
\vspace{-.3cm}
\section{Backhaul Activity - KU Scenario} Knowledge of the active backhaul links may not be always available in many practical scenarios.
In such scenarios, there is no guarantee that the selected transmitter will be active. Hence, in this section, the transmitter selection is performed without the knowledge of the active backhaul links. 
\vspace{-.3cm}
\subsection{Sub-optimal Selection (SS)} 
Using the $\text{S}_k - \text{D}$ link CSI for all ${k \in \{1, 2,\ldots, K\}}$, the SS scheme can be implemented by selecting the transmitter which provides the maximum SNR at the destination, i.e.,
\begin{align} 
\label{eq_gain_max_KU}
k^*=\arg\max_{{k \in \{1, 2,\ldots, K\}}}\{A_D ||\boldsymbol{h}_{Dk}||^2\}.
\end{align}
Including backhaul uncertainty in the KU scenario following \cite{Chinmoy_TVT_2019}, the PDF of the SNR at D can be written as a mixture distribution \footnote[1]{Note that in (6), $\zeta$ denotes the probability that the backhaul link of the \emph{selected} (optimal) transmitter is active.}  as 
\begin{align}\label{PDF_D_backahul_unknown}
 f_{\hat{\gamma}_{D^*}}(x) &= \left(1-\zeta\right)\delta(x) +\zeta f_{\gamma_{D^*}}(x), 
\end{align}
where $\hat{\gamma}_{D^*}$ is the SNR at D corresponding to the transmitter $k^*$ including backhaul uncertainty, $f_{\gamma_{D^*}}(x)$ is the PDF of the SNR $\gamma_{D^*}$ at D if $b_{k^*}$ was active due to a particular selection scheme, and $\delta(x)$ is the delta function. The corresponding CDF is
\begin{align}\label{CDF_D_backahul_unknown}
 F_{\hat{\gamma}_{D^*}}(x) &= \left(1-\zeta\right) +\zeta F_{\gamma_{D^*}}(x).
\end{align}
The CDF $F_{\gamma_{D^*}}(x)$ is obtained as
\begin{align}\label{CDF_D_SUBOP}
 & F_{\gamma_{D^*}}(x)
= \prod_{k=1}^{K}F_{\gamma_{Dk}}(x)=  \left[\frac{\gamma(M,x/A_D)}{\Gamma(M)}\right]^K.
\end{align}

\subsubsection{Derivation of the SOP}
\label{subsec_sopderivation}
The SOP of the system can be derived with the help of (\ref{SOP_SUBOP_BH_Unknown1_1}). Conditioned on the selected transmitter $\text{D}^*$, E experiences its SNR as if there were no backhaul, hence $\hat{\gamma}_{E^*}={\gamma}_{Ek}$ in (\ref{SOP_SUBOP_BH_Unknown1_1}), and its distribution can be obtained following (\ref{pdf_gen}) for $\gamma_{Ek}$. After substitution of \eqref{CDF_D_SUBOP} in \eqref{CDF_D_backahul_unknown} and using \eqref{SOP_SUBOP_BH_Unknown1_1}, we can determine $P_{\text{out}}{(R_{\text{th}})}$ for the SS as
\begin{align}\label{SOP_SUBOP_BH_Unknown2}
 P_{\text{out}}{(R_{\text{th}})} &= (1-\zeta) + \zeta \int_{0}^{\infty}F_{\gamma_{D^*}}(\lambda) f_{\gamma_{Ek}}(y)dy.
\end{align}
 The integral in the above expression can be expressed as
\begin{align}\label{eq_SOPSS_BHU_Nakagami}
&\int_{0}^{\infty}F_{\gamma_{D^*}}((1+y)\rho-1) f_{\gamma_{Ek}}(y)dy\nn\\
&=\frac{1}{
\left(\Gamma(M)\right)^K\Gamma(N)(A_E)^N} \nn\\
& \times \int_{0}^{\infty}\left[\gamma\left(M, \frac{\left(\rho\left(1+y\right)-1\right)}{A_D}\right) \right]^K {y}^{N-1} e^{-\frac{y}{A_E}} dy. \nn\\
\end{align}
The solution involving the incomplete Gamma function does not lead to a simplified expression. However, as $M$ is an integer, by using the lower incomplete Gamma function in this special scenario \cite[eqn. 8.352.6]{ryzhik_2007,Kim_2015_CPSC}, we may write
\begin{align} 
\label{CDF_gen}
 F_{\gamma_{Dk}}(x) =1-e^{-\frac{x}{A_D}}
 \sum_{m=0}^{M-1}\frac{1}{m!}\left(\frac{x}{A_D}\right)^m.
\end{align}
With the help of (\ref{CDF_gen}) and doing some nontrivial mathematical manipulations, the integral in \eqref{SOP_SUBOP_BH_Unknown2} can be expressed as
\begin{align}\label{eq_int_SOPSS_perfectBH_solution}
&\int_{0}^{\infty}F_{\gamma_{D^*}}(\lambda) f_{\gamma_{Ek}}(y)dy\nn\\
&= 1 + \frac{1}{\Gamma(N)(A_E)^N}\sum_{k=1}^{K}{\binom{K}{k}}(-1)^ke^{-\frac{k(\rho-1)}{A_D}} \nonumber \\
 &\times \Upsilon\left(\prod_{m=0}^{M-1}\left(\frac{1}{m!}\right)^{k_m}\right)\left(\frac{1}{A_D}\right)^{\beta_1}\sum_{q=0}^{\beta_1}{\binom{\beta_1}{q}}\nonumber \\
 &\times \left(\rho-1\right)^{\beta_1-q}\rho^{q}\frac{\Gamma(N+q)}{\left(\frac{k\rho}{A_D} +\frac{1}{A_E}\right)^{N+q}},
\end{align}
where $\beta_1 =\sum_{m=0}^{M-1}mk_m$ and 
\begin{align}
\Upsilon \triangleq \sum_{k_0+k_1+...+k_{M-1}=k} \left(\frac{k!}{k_0! k_1!,...,k_{\left(M-1\right)}!}\right).
\end{align}
The summation $\sum_{k_0+k_1+...+k_{M-1}=k}$ represents the sum over all the possible values of $k_m$, $k_m\in\{0,1,\ldots,k\}$, such that $k_0+k_1+...+k_{M-1}=k$, which is a result of the multinomial expansion of 
$\left(\sum_{m=0}^{M-1}\frac{1}{m!}x^m\right)^{k}$ in (\ref{CDF_gen}), and the summation over $q$ comes from the binomial expansion of $\lambda^{\beta_1}$.
$P_{\text{out}}{(R_{\text{th}})} $ can be obtained after substituting \eqref{eq_int_SOPSS_perfectBH_solution} in \eqref{SOP_SUBOP_BH_Unknown2}.
Note that the SS rule only requires transmitters to have destination channel knowledge, which may reduce feedback overhead and complexity. Furthermore, eavesdropper channel knowledge may not be available. 

\subsection{Optimal Selection (OS)}
 In optimal selection, we select the transmitter for which the secrecy capacity of the system is maximized without knowing which backhaul links are active, i.e., 
 \begin{align} 
\label{eq_optimal_max}
k^*=\arg\max_{k \in \{1, 2,\ldots, K\}}\{C_{S,k}\},
\end{align}
 where $C_{S,k}$ is the secrecy capacity for the individual $k$th transmitter. Due to the independent and identically distributed links, the SOP including backhaul reliability in the KU scenario as in (\ref{SOP_SUBOP_BH_Unknown2}) is obtained as
 \begin{align}\label{eq_SOP_OS}
P_{\text{out}}{(R_{\text{th}})} & =(1-\zeta)+\zeta\mathbb{P}\lb[\max_{{k \in \{1, 2,\ldots, K\}}}\{C_{S,k}\}< R_{\text{th}}\rb] \nn\\
 &=(1-\zeta)+\zeta\lb(\mathbb {P}\lb[C_{S,k}< R_{\text{th}}\rb]\rb)^K.
 \end{align}
We can observe from (\ref{eq_SOP_OS}) that the derivation of the optimal SOP with backhaul uncertainty requires the SOP of the individual transmitters with active backhaul links. The SOP of an individual transmitter can be obtained following (\ref{SOP_SUBOP_BH_Unknown1_1}) in the case of active backhaul links as

\begin{align}\label{SOP_K=1_BH_Unknown}
  &\mathbb{P}[C_{S,k}< R_{\text{th}}] 
 =\int_{0}^{\infty}{F}_{\gamma_{Dk}}(\lambda) f_{\gamma_{Ek}}(y)d{y}\nonumber \\
 &=1-\frac{ e^{-\frac{(\rho-1)}{A_D}}}{\Gamma(N)A_E^N}\sum_{m=0}^{M-1} \left(\frac{1}{m!}\right)\left(\frac{1}{A_D}\right)^{m}\nonumber\\
 &\times \sum_{q=0}^{m}\left(\rho-1\right)^{m-q}\rho^{q}\frac{\Gamma(N+q)}{\left(\frac{\rho}{A_D} +\frac{1}{A_E}\right)^{N+q}}. 
\end{align}
 Inserting (\ref{SOP_K=1_BH_Unknown}) into (\ref{eq_SOP_OS}), the optimal SOP is obtained.

The OS selection scheme requires global CSI knowledge (except for backhaul knowledge in the KU scenario), which leads to an increased feedback overhead and complexity over the SS scheme. However, the performance of the OS scheme may also serve as a (best-case) performance bound for the SS scheme.
\vspace{-.3cm}
\section{ Backhaul Activity - KA Scenario}
When the knowledge of active backhaul is available \textit{a priori}, the selection can be performed within the set of transmitters whose backhaul is active. This is expected to improve the secrecy performance over the KU scenario. In this scenario, the SNR distribution corresponding to each individual transmitter including backhaul uncertainty  is 
expressed following \cite{Kim_2016_CPSC_Globecom} as a mixture distribution\footnote[2]{Note that in contrast to (6), here $\zeta$ denotes the probability that the backhaul link of the $k$-th transmitter is active. The proposed mixture distribution in (6) and (17) help us to find the performance in the KU and KA scenarios in a unified manner.} as
\vspace{-.1cm}
\begin{align}\label{pdf_D}
 f_{\hat{\gamma}_{Dk}}(x) &= \left(1-\zeta\right)\delta(x) +\zeta f_{{\gamma_{Dk}}}(x). 
\end{align}
Accordingly, the CDF is derived as
\begin{align}\label{CDF_D}
 F_{\hat{\gamma}_{Dk}}(x) =1-\zeta e^{\frac{-x}{A_D}}
 \sum_{m=0}^{M-1}\frac{1}{m!}\left(\frac{x}{A_D}\right)^m.
\end{align}
\subsection{Sub-optimal Selection (SS)} 
In SS, the transmitter is selected from the set of transmitters $\mathcal{S}$ with active backhaul as
\begin{align} 
\label{eq_gain_max_KA}
k^*=\arg\max_{k \in \mathcal{S}}\{A_D ||\boldsymbol{h}_{Dk}||^2\}.
\end{align}
The CDF of the received SNR at D including backhaul uncertainty can be written as
\begin{align}\label{eqn3}
&F_{\hat{\gamma}_{D^*}}(x)= \prod_{k=1}^{K}\mathbb{P}\left[\hat{\gamma}_{Dk} \leq x \right] = \left[F_{\hat{\gamma}_{D}}(x) \right]^K \nonumber\\
&=1 +\sum_{k=1}^{K}{\binom{K}{k}}(-\zeta)^k e^{-\frac{kx}{A_D}}\left(\sum_{m=0}^{M-1}\frac{1}{m!}\left(\frac{x}{A_D}\right)^m\right)^{k}.
\end{align}
Here $\hat{\gamma}_{E^*}={\gamma}_{Ek}$ as in (\ref{SOP_SUBOP_BH_Unknown2}), due to the same reasoning.
Using a methodology similar to that of subsection \ref{subsec_sopderivation}, using (\ref{eqn3}) and following (\ref{pdf_gen}) the SOP for the SS scheme in the KA scenario is obtained as
\begin{align}\label{SOP_SUBOP_GEN}
 & P_{\text{out}}{(R_{\text{th}})}= \mathbb{P}\lb[\frac{1+\hat{\gamma}_{D^*}}{1+\hat{\gamma}_{E^*}} < \rho\rb]\nonumber \\
 &= 1 + \frac{1}{\Gamma(N)(A_E)^N}\sum_{k=1}^{K}{\binom{K}{k}}(-\zeta)^ke^{-\frac{k(\rho-1)}{A_D}} \nonumber \\
 &\times \Upsilon\left(\prod_{m=0}^{M-1}\left(\frac{1}{m!}\right)^{k_m}\right)\left(\frac{1}{A_D}\right)^{\beta_1}\sum_{q=0}^{\beta_1}{\binom{\beta_1}{q}}\nonumber \\
 &\times \left(\rho-1\right)^{\beta_1-q}\rho^{q}\frac{\Gamma(N+q)}{\left(\frac{k\rho}{A_D} +\frac{1}{A_E}\right)^{N+q}}.
\end{align}

\subsection{Optimal Selection (OS)} 
Optimal selection is performed using the definition of the selected transmitter as in (\ref{eq_optimal_max}); however, this time the selection must be from the set of transmitters with active backhaul links, $\mathcal{S}$.
The SOP in this scenario is obtained using $F_{\hat{\gamma}_{Dk}}(\cdot)$ from (\ref{CDF_D}), considering $\hat{\gamma}_{Ek}={\gamma}_{Ek}$, taking $f_{{\gamma}_{Ek}}(\cdot)$ from (\ref{pdf_gen}) and with the help of (\ref{SOP_K=1_BH_Unknown}) as
\begin{align}\label{OP_SEL}
 &P_{\text{out}}{(R_{\text{th}})}=\lb(\mathbb{P}\lb[C_{S,k}< R_{\text{th}}\rb]\rb)^K \nonumber \\
 &=\left(\int_{0}^{\infty} F_{\hat{\gamma}_{Dk}}(\lambda) f_{\hat{\gamma}_{Ek}}(y)dy\right)^K\nonumber\\
 &=\lb[1-\frac{\zeta e^{-\frac{(\rho-1)}{A_D}}}{\Gamma(N)A_E^N}\sum_{m=0}^{M-1} \left(\frac{1}{m!}\right)\left(\frac{1}{A_D}\right)^{m}\rb.\nonumber \\ &\lb.\mathsf{x}\sum_{q=0}^{m}\left(\rho-1\right)^{m-q}\rho^{q}\frac{\Gamma(N+q)}{\left(\frac{\rho}{A_D} +\frac{1}{A_E}\right)^{N+q}} \rb]^K. 
\end{align}

\textit{Remark 1:} The application of the mixture distribution in the form of (\ref{PDF_D_backahul_unknown}) and (\ref{pdf_D}) in the KU and KA scenario, respectively, has enabled us to analyze both the backhaul activity knowledge scenarios in a generalized way. The proposed analysis method helps us to find the performance for the SS and OS schemes in the KU and KA scenarios in a unified manner.

\section{Asymptotic Analysis}
To understand the asymptotic behaviour at high SNR, we set $P_T/{\sigma^2}\rightarrow\infty$ and find expressions for the SOP of all the selection schemes proposed in the previous section in both KU and KA scenarios.
\subsection{KU scenario}
 \subsubsection{SS with KU}
To understand the behaviour of \eqref{SOP_SUBOP_BH_Unknown2} as $P_T/{\sigma^2}$ increases, we transform \eqref{SOP_SUBOP_BH_Unknown2} as a function of $P_T/{\sigma^2}$ as
\begin{align}\label{deriv_sun_op_asymp_bh_un1}
 &\underset{P_T/{\sigma^2}\rightarrow\infty}{P_{\text{out}}{(R_{\text{th}})}}
 =1+ \frac{\zeta a^N}{\Gamma(N)}\sum_{k=1}^{K}{\binom{K}{k}}(-1)^k \nonumber\\
 &\times e^{-\frac{k(\rho-1)}{a P_T/\sigma^2}} \Upsilon \left(\prod_{m=0}^{M-1}\left(\frac{1}{m!}\right)^{k_m}\right)\left(\frac{1}{a}\right)^{\beta_1}\nonumber \\
  &\times \sum_{q=0}^{\beta_1}\left(\frac{1}{P_T/\sigma^2}\right)^{\beta_1 -q}{\binom{\beta_1}{q}}\left(\rho-1\right)^{\beta_1-q}\nonumber \\
  &\times(\rho a b)^{q}\frac{\Gamma(N+q)}{\left(k\rho b +a\right)^{N+q}}. 
\end{align}
We observe that as $P_T/\sigma^2 \rightarrow \infty$ in the above equation, $\exp\lb(-{\frac{k(\rho-1)}{a P_T/\sigma^2}}\rb) \rightarrow 1$ and the summation terms for $q=0$ to $\beta_1 -1$ will tend to zero due to the condition $\left(\frac{1}{P_T/\sigma^2}\right)^{\beta_1-q}\rightarrow 0$. The only nonzero term will correspond to the term for $q=\beta_1$, which is found as
\begin{align}\label{deriv_sun_op_asymp_bh_un2}
 &\underset{P_T/{\sigma^2}\rightarrow\infty}{P_{\text{out}}{(R_{\text{th}})}}= 1 +\frac{\zeta a^N}{\Gamma(N)}\sum_{k=1}^{K}{\binom{K}{k}}(-1)^k\nonumber\\
 &\times \Upsilon\left(\prod_{m=0}^{M-1}\left(\frac{1}{m!}\right)^{k_m}\right)(\rho b)^{\beta_1} \frac{\Gamma(N+\beta_1)}{\left(k\rho b +a\right)^{N+\beta_1}}.
\end{align}

\textit{Remark 2:} It can be observed from (\ref{deriv_sun_op_asymp_bh_un2}) that the SOP is independent of the transmitted power $P_T$ but depends on the number of transmitters $K$ and all of the channel parameters. This implies that the SOP saturates to a constant value which is determined by the number of transmitters $K$ and all of the channel parameters. 

\subsubsection{S with KU}
Similarly for optimal selection, the asymptotic analysis of SOP can be performed by representing \eqref{eq_SOP_OS} in terms of $P_T/{\sigma^2}$ and then proceeding as per \eqref{deriv_sun_op_asymp_bh_un1}, yielding
\begin{align}\label{asymp_os_BHU}
 & \underset{P_T/{\sigma^2}\rightarrow\infty}{P_{\text{out}}{(R_{\text{th}})}}=
 1+\zeta\sum_{k=0}^{K}{\binom{K}{k}}(-1)^k\left(\frac{a^N}{\Gamma(N)}\right)^k \nonumber\\
 &\times \Upsilon \left(\prod_{m=1}^{M-1} \left(\frac{1}{m!}\right)^{k_m} \right)(\rho b)^{\beta_1} \left(\frac{\Gamma(N+m)}{\left(\rho b +a\right)^{N+m}}\right)^{\beta_2},
\end{align}
where $\beta_2 =\sum_{m=0}^{M-1}k_m$.
\textit{Remark 3}: Following (\ref{deriv_sun_op_asymp_bh_un2}) and (\ref{asymp_os_BHU}), we observe that in contrast to the KU scenario, the term $\zeta$ is outside the second term. Hence, its contribution in asymptotic limit of SOP is independent of the number of transmitters.
\subsection{KA scenario}
 \subsubsection{SS with KA}
The asymptotic analysis of the SOP for the SS scheme can be obtained from 
\eqref{SOP_SUBOP_GEN} by substituting $P_T/{\sigma^2}\rightarrow\infty$ in a similar manner to (\ref{deriv_sun_op_asymp_bh_un1}); this yields 
\begin{align}\label{sub_op_asymp_BH}
 &\underset{P_T/{\sigma^2}\rightarrow\infty}{P_{\text{out}}{(R_{\text{th}})}}=1 +\frac{a^N}{\Gamma(N)}\sum_{k=1}^{K}{\binom{K}{k}}(-1)^k(\zeta)^k \nonumber \\
 \nonumber \\
 &\times \Upsilon \left(\prod_{m=0}^{M-1}\left(\frac{1}{m!}\right)^{k_m}\right)(\rho b)^{\beta_1} \frac{\Gamma(N+\beta_1)}{\left(k\rho b +a\right)^{N+\beta_1}}. 
\end{align}
\subsubsection{S with KA}
The asymptotic analysis of the SOP in OS scheme can be obtained by representing \eqref{OP_SEL} in terms of $P_T/{\sigma^2}$ and proceeding as per \eqref{deriv_sun_op_asymp_bh_un1}, yielding 
\begin{align}\label{op_asym_BH}
 &\underset{P_T/{\sigma^2}\rightarrow\infty}{P_{\text{out}}{(R_{\text{th}})}}=
 1+\sum_{k=0}^{K}{\binom{K}{k}}(-1)^k(\zeta)^k\left(\frac{a^N}{\Gamma(N)}\right)^k\nn\\
 &\times\Upsilon \left(\prod_{m=1}^{M-1} \left(\frac{1}{m!}\right)^{k_m} \right)(\rho b)^{\beta_1} \left(\frac{\Gamma(N+m)}{\left(\rho b +a\right)^{N+m}}\right)^{\beta_2}.
\end{align}

\textit{Remark 4}: From (\ref{sub_op_asymp_BH}) and (\ref{op_asym_BH}), we observe that in the KA scenario, the term containing $\zeta$ is \textit{inside} the summation. Hence, as the number of transmitters increases, since $\zeta<1$,  the higher-order terms in \eqref{sub_op_asymp_BH} and \eqref{op_asym_BH} decrease significantly in contrast to the corresponding terms in \eqref{deriv_sun_op_asymp_bh_un2} and \eqref{asymp_os_BHU}, respectively. Therefore the asymptotic SOP becomes more sensitive to the number of transmitters compared to the KU scenario, especially for lower values of the backhaul reliability factor.
\vspace{-.1cm}
\section{Results}
\label{sec_results_and_discussions}
Numerical and simulation results are presented in this section to validate the derived expressions. Results are plotted against the average SNR $P_T/\sigma^2$. Unless otherwise specified, the following parameters are considered for the numerical results. The path loss factors for D and E links are assumed to be $a=0.5$ and $b=0.2$, respectively. Further, the number of multipath components in the D and E links are taken as $M=6$ and $N=4$, respectively. The rate threshold, $R_{\text{th}}$, is set to $1$ for the analysis. The plots for optimal and sub-optimal schemes are marked OS and SS, respectively. In all graphs, the analytical curves are represented by lines with markers and the corresponding simulation curves by only black colored `x' markers. It can be observed that all of the simulation curves match with the analytical curves, thus validating our analysis. 
Further, comparing the OS and SS schemes for any given set of parameters, it is clear that OS performs better than SS but at the cost of additional computational complexity and signaling overhead.


\begin{figure}
\vspace{-.1cm}
 \centering
\includegraphics[width=3.5in]{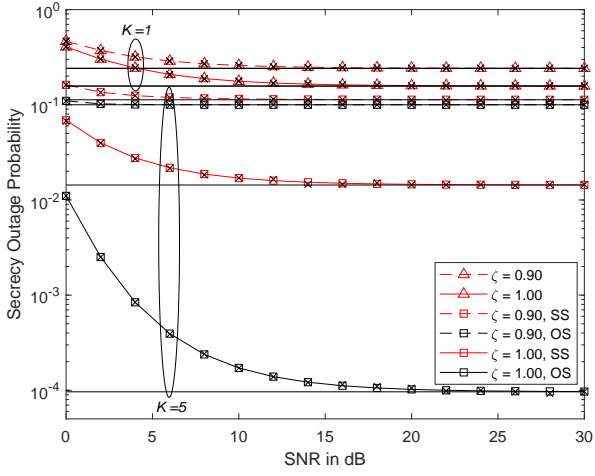} 
 \vspace{-.3cm}
 \caption{SOP vs. SNR in KU scenario for variation in $K$, $\zeta$ and selection scheme. }
 \label{fig2_1_SOP_VS_SNR_for_K_ZETA_STS_OTS_BH_KU}
 \vspace{-.6cm}
 \end{figure}
 Fig. \ref{fig2_1_SOP_VS_SNR_for_K_ZETA_STS_OTS_BH_KU} and Fig. \ref{fig2_2_SOP_VS_SNR_for_K_ZETA_STS_OTS_BH_KA} show the effect of $K$, $\zeta$, and the knowledge of backhaul availability on the SOP for the given parameter set of $M=6$, $N=4$, $a=0.5$, $b=0.2$, and $R_{\text{th}}=1$. Fig. \ref{fig2_1_SOP_VS_SNR_for_K_ZETA_STS_OTS_BH_KU} represents the KU scenario while  Fig. \ref{fig2_2_SOP_VS_SNR_for_K_ZETA_STS_OTS_BH_KA} is for KA scenario. The black colored horizontal lines represent the asymptotic limits of the corresponding SOP performance. In general, we can observe that the SOP performance improves as $K$ increases due to the fact that diversity gain is achieved at the destination in the presence of multiple transmitters. However, by comparing Fig. \ref{fig2_1_SOP_VS_SNR_for_K_ZETA_STS_OTS_BH_KU} and Fig. \ref{fig2_2_SOP_VS_SNR_for_K_ZETA_STS_OTS_BH_KA} we find that the performance improvement with the with increasing $K$ is higher in the KA scenario. This is in agreement with Remark 4. It is also confirmed that lower value of backhaul reliability factor degrades the secrecy performance for both the scenarios. Furthermore, it is observed that the \textit{a priori} knowledge of backhaul activity always improves the SOP performance.
 \begin{figure}
 \vspace{-.1cm}
 \centering
\includegraphics[width=3.5in]{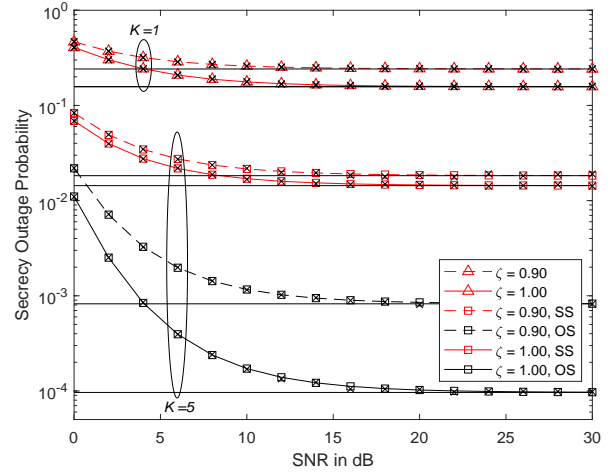} 
 \vspace{-.3cm}
 \caption{SOP vs. SNR in KA scenario for variation in $K$, $\zeta$ and selection scheme. }
 \label{fig2_2_SOP_VS_SNR_for_K_ZETA_STS_OTS_BH_KA}
 \vspace{-.6cm}
 \end{figure}
 
\begin{figure}
\vspace{-.1cm}
 \centering
\includegraphics[width=3.5in]{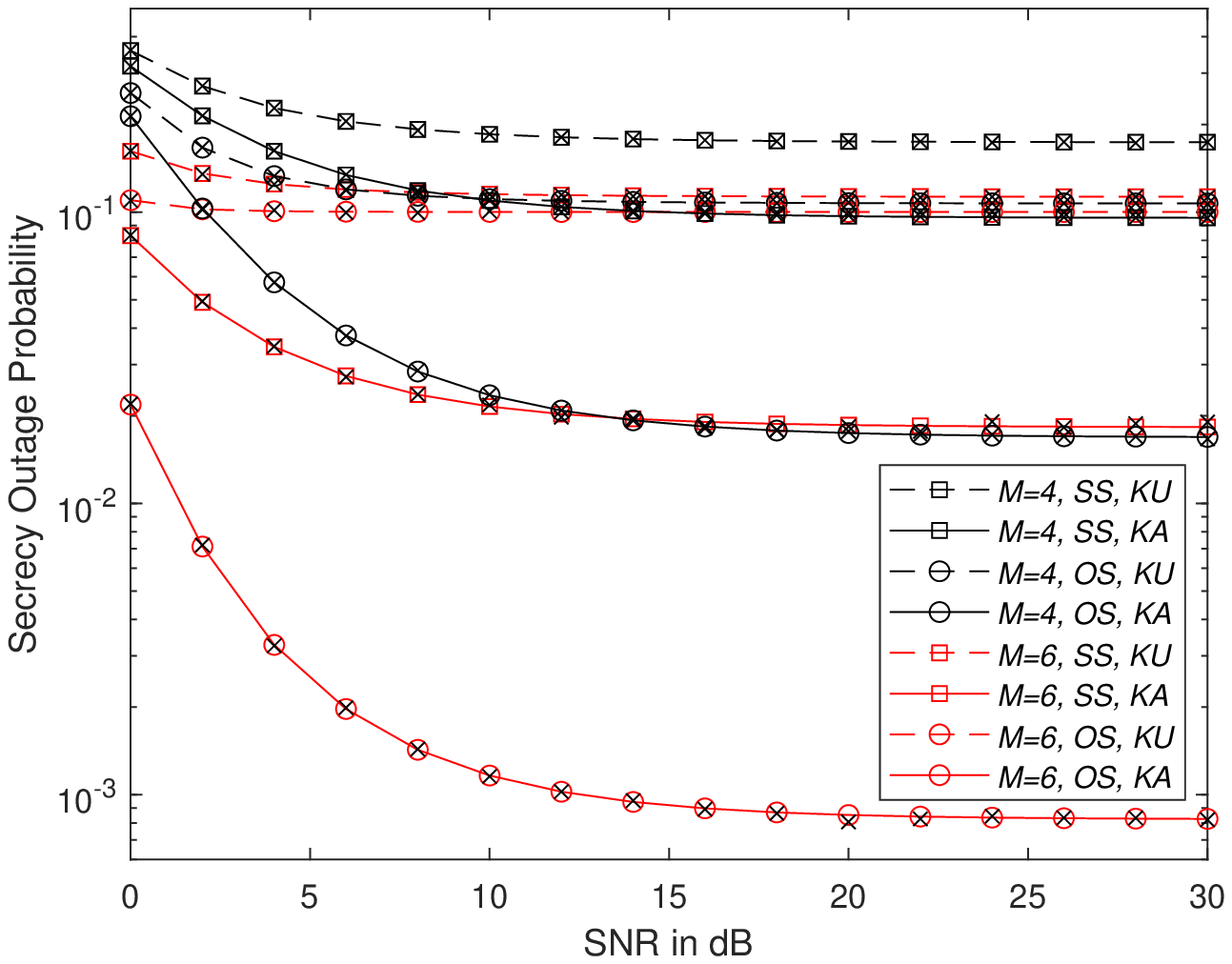} 
 \vspace{-.3cm}
 \caption{SOP vs. SNR for variation in $M$ and selection scheme in KU and KA Scenario. $N$ is fixed at 4. }
 \label{fig3_SOP_VS_SNR_for_MULTIPATH_STS_OTS_VARIATION_IN_M}
\vspace{-.6cm}
 \end{figure}
 
  \begin{figure}
\vspace{-.1cm}
 \centering
\includegraphics[width=3.5in]{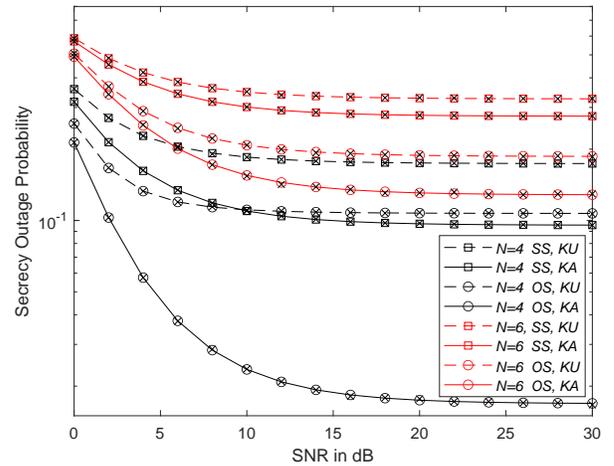} 
 \vspace{-.3cm}
 \caption{SOP vs. SNR for variation in $N$ and selection scheme in KU and KA Scenario. $M$ is fixed at 4. }
 \label{fig4_SOP_VS_SNR_for_MULTIPATH_STS_OTS_VARIATION_IN_N}
 \vspace{-.6cm}
 \end{figure}
 
Fig. \ref{fig3_SOP_VS_SNR_for_MULTIPATH_STS_OTS_VARIATION_IN_M} and Fig. \ref{fig4_SOP_VS_SNR_for_MULTIPATH_STS_OTS_VARIATION_IN_N} show the effect of changing the number of channel paths ($M$ and $N$) on the SOP performance for the OS and SS schemes in both KU and KA scenarios. The common parameters set for these two figures are $K=5$, $a=0.5$, $b=0.2$, and $\zeta=0.9$. From the SOP curves in Fig.  \ref{fig3_SOP_VS_SNR_for_MULTIPATH_STS_OTS_VARIATION_IN_M}, it is observed that by keeping number of paths in the eavesdropper channel $(N)$ constant, an increase in the number of paths in the destination channel $(M)$ improves the secrecy performance for the SS as well as the OS schemes. This effect can be observed in the KU as well as KA scenario. However, by keeping $M$ constant and increasing $N$ has the opposite effect, as can be observed in Fig. \ref{fig4_SOP_VS_SNR_for_MULTIPATH_STS_OTS_VARIATION_IN_N}. This can be attributed to the fact that the use of SC-CP improves the SNR corresponding to that channel when the number of paths increases, without any modification at the receiver. It can also be observed from Fig. \ref{fig3_SOP_VS_SNR_for_MULTIPATH_STS_OTS_VARIATION_IN_M} and Fig. \ref{fig4_SOP_VS_SNR_for_MULTIPATH_STS_OTS_VARIATION_IN_N} that knowledge of active backhaul links plays an important role in improving the SOP performance, and for the same set of parameters, OS outperforms SS.
 
 \begin{figure}
  \vspace{-.3cm}
 \centering
\includegraphics[width=3.5in]{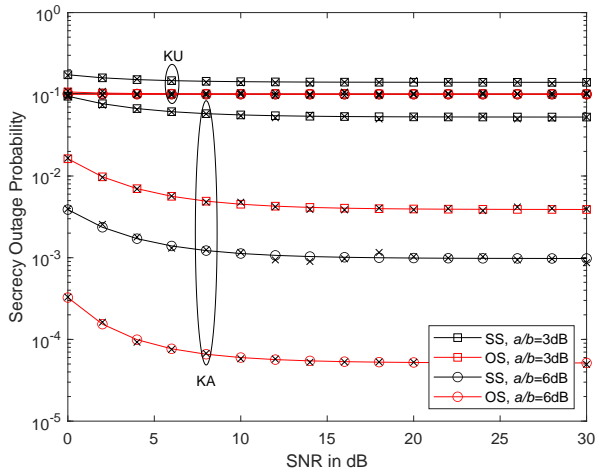} 
 \vspace{-.3cm}
 \caption{SOP vs. SNR for variation in path loss and selection scheme. }
 \label{fig5_SOP_VS_SNR_for_PATHLOSS_STS_OTS_BH_VARIATION}
 \vspace{-.6cm}
 \end{figure}
Fig. \ref{fig5_SOP_VS_SNR_for_PATHLOSS_STS_OTS_BH_VARIATION} shows the impact of the ratio $a/b$ on the SOP performance for SS and OS schemes in the KU and KA scenarios with the parameter set $K=5$, $M=6$, $N=4$, and $\zeta=0.9$. We observe that the SOP performance improves with increasing values of $a/b$. This observation can be justified by the fact that a higher value of $a/b$ represents a more favorable channel to the destination in comparison to the eavesdropper. Further, we observe that the effect of $a/b$ parameter variation has strong impact on the SOP performance when the backhaul reliability knowledge is available before selection as compared to the scenario when it is not. This observation is true for both OS and SS schemes.
\vspace{-.1cm}
\section*{Acknowledgement}This publication has emanated from research conducted with the financial support of Science Foundation Ireland (SFI) under Grant Number 17/US/3445 and Science and Engineering Research Board India
sponsored project under Grant Number ECR/2018/002795. 
\section{Conclusions}
In this paper, we presented the secrecy performance analysis for optimal and sub-optimal transmitter selection schemes with or without the knowledge of the active backhaul links for SC-CP modulation over frequency selective fading channels. Closed-form expressions for the SOP is presented along with its asymptotic analysis. We showed that at high SNR, the secrecy performance approaches a fixed value governed by multiple parameters including the backhaul link reliability, number of transmitters, number of channel paths, pathloss factors and secrecy rate threshold. We observed that knowledge of the active backhaul links significantly improved the secrecy performance. It was also shown that the OS rule benefited the most from the active backhaul knowledge; however, on the other hand SS can reduce feedback overhead and system complexity. 
\vspace{-.2cm}
\bibliographystyle{IEEEtran}
\vspace{-.2cm}
\bibliography{IEEEabrv,ref}

\begin{thebibliography}{10}
\providecommand{\url}[1]{#1}
\csname url@samestyle\endcsname
\providecommand{\newblock}{\relax}
\providecommand{\bibinfo}[2]{#2}
\providecommand{\BIBentrySTDinterwordspacing}{\spaceskip=0pt\relax}
\providecommand{\BIBentryALTinterwordstretchfactor}{4}
\providecommand{\BIBentryALTinterwordspacing}{\spaceskip=\fontdimen2\font plus
\BIBentryALTinterwordstretchfactor\fontdimen3\font minus
  \fontdimen4\font\relax}
\providecommand{\BIBforeignlanguage}[2]{{%
\expandafter\ifx\csname l@#1\endcsname\relax
\typeout{** WARNING: IEEEtran.bst: No hyphenation pattern has been}%
\typeout{** loaded for the language `#1'. Using the pattern for}%
\typeout{** the default language instead.}%
\else
\language=\csname l@#1\endcsname
\fi
#2}}
\providecommand{\BIBdecl}{\relax}
\BIBdecl

\bibitem{Leung_1978_Gaussian_Wiretap_Channel}
S.~{Leung-Yan-Cheong} and M.~{Hellman}, ``The {Gaussian} wire-tap channel,''
  \emph{IEEE Transactions on Information Theory}, vol.~24, no.~4, pp. 451--456,
  Jul. 1978.

\bibitem{Mukherjee_2011_beamforming}
A.~Mukherjee and A.~L. Swindlehurst, ``Robust beamforming for security in
  {MIMO} wiretap channels with imperfect {CSI},'' \emph{IEEE Transactions on
  Signal Processing}, vol.~59, no.~1, pp. 351--361, Jan. 2011.

\bibitem{Sheng_2018_beaforming_difficult}
Z.~Sheng, H.~D. Tuan, T.~Q. Duong, and H.~V. Poor, ``Beamforming optimization
  for physical layer security in {MISO} wireless networks,'' \emph{IEEE
  Transactions on Signal Processing}, vol.~66, no.~14, pp. 3710--3723, Jul.
  2018.

\bibitem{Chinmoy_TWC_2015}
C.~{Kundu}, S.~{Ghose}, and R.~{Bose}, ``Secrecy outage of dual-hop
  regenerative multi-relay system with relay selection,'' \emph{IEEE
  Transactions on Wireless Communications}, vol.~14, no.~8, pp. 4614--4625,
  Aug. 2015.

\bibitem{Chinmoy_GC16}
C.~{Kundu}, T.~M.~N. {Ngatched}, and O.~A. {Dobre}, ``Relay selection to
  improve secrecy in cooperative threshold decode-and-forward relaying,'' in
  \emph{Proc. IEEE Global Communications Conference}, Washington, DC, USA, Dec.
  2016, pp. 1--6.

\bibitem{Chinmoy_GC17}
S.~{Ghose}, C.~{Kundu}, and O.~A. {Dobre}, ``Secrecy outage of proactive relay
  selection by eavesdropper,'' in \emph{Proc. Global Communications
  Conference}, Singapore, Dec. 2017, pp. 1--6.

\bibitem{Kim_2015_CPSC}
L.~{Wang}, K.~J. {Kim}, T.~Q. {Duong}, M.~{Elkashlan}, and H.~V. {Poor},
  ``Security enhancement of cooperative single carrier systems,'' \emph{IEEE
  Transactions on Information Forensics and Security}, vol.~10, no.~1, pp.
  90--103, Jan. 2015.

\bibitem{other_2017_CPSC}
H.~T. {Nguyen}, J.~{Zhang}, N.~{Yang}, T.~Q. {Duong}, and W.~{Hwang}, ``Secure
  cooperative single carrier systems under unreliable backhaul and dense
  networks impact,'' \emph{IEEE Access}, vol.~5, pp. 18\,310--18\,324, Jul.
  2017.

\bibitem{Kim_2016_CPSC_Globecom}
P.~L. {Yeoh}, K.~J. {Kim}, P.~V. {Orlik}, and H.~V. {Poor}, ``Secrecy
  performance of cooperative single carrier systems with unreliable backhaul
  connections,'' in \emph{Proc. IEEE Global Communications Conference},
  Washington, DC, USA, Dec. 2016, pp. 1--6.

\bibitem{Kim_2016_CPSC_Trans}
K.~J. {Kim}, P.~L. {Yeoh}, P.~V. {Orlik}, and H.~V. {Poor}, ``Secrecy
  performance of finite-sized cooperative single carrier systems with
  unreliable backhaul connections,'' \emph{IEEE Transactions on Signal
  Processing}, vol.~64, no.~17, pp. 4403--4416, Sep. 2016.

\bibitem{Kim_2018_CDD_JOUR}
H.~{Liu}, P.~L. {Yeoh}, K.~J. {Kim}, P.~V. {Orlik}, and H.~V. {Poor}, ``Secrecy
  performance of finite-sized in-band selective relaying systems with
  unreliable backhaul and cooperative eavesdroppers,'' \emph{IEEE Journal on
  Selected Areas in Communications}, vol.~36, no.~7, pp. 1499--1516, Jul.
  2018\color{black}.

\bibitem{Chinmoy_TVT_2019}
J.~{Zhang}, C.~{Kundu}, O.~A. {Dobre}, E.~{Garcia-Palacios}, and N.~{Vo},
  ``Secrecy performance of small-cell networks with transmitter selection and
  unreliable backhaul under spectrum sharing environment,'' \emph{IEEE
  Transactions on Vehicular Technology}, vol.~68, no.~11, pp. 10\,895--10\,908,
  Nov. 2019.

\bibitem{Shalini_GC20}
S.~{Tripathi}, C.~{Kundu}, O.~A. {Dobre}, A.{Bansal}, and M.~F. {Flanagan},
  ``Recurrent neural network assisted transmitter selection for secrecy in
  cognitive radio network,'' in \emph{Proc. IEEE Global Communications
  Conference}, Taipei, Taiwan, Dec. 2020, pp. 1--6.

\bibitem{backhaul_Yin}
C.~{Yin}, E.~{Garcia-Palacios}, N.~{Vo}, and T.~Q. {Duong}, ``Cognitive
  heterogeneous networks with multiple primary users and unreliable backhaul
  connections,'' \emph{IEEE Access}, vol.~7, pp. 3644--3655, Dec. 2018.

\bibitem{bailleu_2020}
J.~{Bailleul}, L.~{Jacobs}, M.~{Guenach}, and M.~{Moeneclaey}, ``Optimized
  precoded spatio-temporal partial-response signaling over frequency-selective
  {MIMO} channels,'' \emph{IEEE Transactions on Wireless Communications},
  vol.~19, no.~9, pp. 5938--5950, Sep. 2020.

\bibitem{Tubbax_SC_CP_first}
J.~{Tubbax}, B.~{Come}, L.~{Van der Perre}, L.~{Deneire}, S.~{Donnay}, and
  M.~{Engels}, ``{OFDM} versus single carrier with cyclic prefix: a
  system-based comparison,'' in \emph{Proc. IEEE Vehicular Technology
  Conference}, Atlantic city, NJ, USA, Oct. 2001, pp. 1115--1119.

\bibitem{ryzhik_2007}
I.~S. Gradshteyn and I.~M. Ryzhik, \emph{Table of Integrals, Series, and
  Products}, 7th~ed.\hskip 1em plus 0.5em minus 0.4em\relax Elsevier/Academic
  Press, Amsterdam, Netherland, 2007.

\end{thebibliography}
\end{document}